\documentclass[11pt,twoside]{article}

\usepackage{baaa}

\usepackage{graphicx}
\usepackage{subfigure}
\usepackage{amssymb}
\usepackage[spanish,activeacute,english]{babel}
\usepackage[latin1]{inputenc}
\usepackage{verbatim}
\usepackage{amsmath}
\usepackage{amsfonts}
\usepackage{amssymb}
\usepackage[colorlinks=true,dvips]{hyperref}
\usepackage{natbib}

\begin{document}
%%%%%%%%%%%%%%%%%%%%%%%%%%%%%%%%%%%%%%%%%%%%%%%%%%%%%%%%%%%%%%%%%%%%%%%%%%
%%%% SELECCIONE EL IDIOMA EN QUE SE ESCRIBE EL ARTÃÂÃÂCULO:              %%%%
%\myselectspanish
\myselectenglish
%%%%%%%%%%%%%%%%%%%%%%%%%%%%%%%%%%%%%%%%%%%%%%%%%%%%%%%%%%%%%%%%%%%%%%%%%%
\vskip 1.0cm
\markboth{Benaglia et al.}%
{Runaway stars and the ISM}

\pagestyle{myheadings}
%%%% DESCOMENTE LA LINEA QUE DESCRIBE EL CARACTER DE SU TRABAJO       %%%%
\vspace*{0.5cm}

\noindent TRABAJO INVITADO 
%\noindent PRESENTACIÃÂÃÂN ORAL
%\noindent PRESENTACIÃÂÃÂN MURAL
%\noindent RESUMEN 

\vskip 0.3cm
\title{Runaway stars: their impact on the intestellar medium}

%\title{ Template paper for publication in the Bulletin of the 
%Argentinian Astronomical Association with instructions for the use of 
%\LaTeX{}}
 
\author{P. Benaglia$^{1,2}$, I. R. Stevens$^{3}$, C. S. Peri$^{1,2}$
% \& D. S. Brookes$^{3}$
}

\affil{%
  (1) Instituto Argentino de Radioastronom\'{\i}a, CCT-La Plata, CONICET\\ 
  (2) Facultad de Ciencias Astron\'omicas y Geof\'{\i}sicas, Universidad Nacional de La Plata\\
  (3) School of Physics and Astronomy, University of Birmingham, Edgbaston, Birmingham, B15 2TT, UK
}

\begin{abstract} 
Runaway, massive stars are not among the most numerous.
However, the bow shocks built by their supersonic movement
in the interstellar medium have been detected in the infrared range in
many cases. Most recently, the stellar bow shocks
have been proposed as particle acceleration sites, as radio
data analysis at high angular resolution have shown. 
We present results of different manifestations of the
stellar bowshock phenomenon, revealed from modern IR
databases.
\end{abstract}

\begin{resumen}
Las estrellas masivas fugitivas no son de las m\'as numerosas. 
Sin embargo, los $bowshocks$ formados debido a su movimiento supers\'onico 
en el medio interestelar han sido detectados en el rango infrarrojo en 
muchos casos. Muy recientemente, estos $bowshocks$ estelares fueron
propuestos como sitios de aceleraci\'on de part\'{\i}culas, como lo sugiere 
el an\'alisis de datos de alta resoluci\'on angular a bajas frecuencias 
de radio. Se presentan aqu\'{\i} resultados de distintas 
manifestaciones relacionadas con $bowshocks$ estelares, revelados a partir 
de las bases de datos IR m\'as modernas. 
\end{resumen}

\section{Runaway stars} 
%\label{}

The seminal studies on stars with high velocities ($v_* \gtrapprox$ tens of km 
s$^{-1}$) were carried out during the '50s by A. Blaauw and collaborators. 
Among other findings, they pointed out that an important number of early-type 
stars (O-B5) move faster than the surrounding objects. The motion of some of 
them could be interpreted as if the stars are escaping from their parental
%-sometimes common to a pair of stars- 
associations: see Blaauw \& Morgan (1954) and the case of the stars AE Aur and 
$\mu$ Col. The name `runaway stars' was thus coined by Blaauw (1961), who, 
together with Zwicky (1957), suggested that the velocity kick could be given by 
a supernova companion in an originally binary system. In those first studies, 
the velocity distribution of OB stars with $v_*$ up to $\sim$ 30 km s$^{-1}$ 
was reasonably fitted by a gaussian, but an important number of stars with 
greater velocities failed the fit. After compiling proper motion information of 
O stars up to 2 kpc, Stone (1979) concluded that there were two stellar 
populations: one of low spatial velocities (below 25 km s$^{-1}$) and the rest 
with higher velocities (see Figure 1). 
He fitted both velocity distributions with gaussian 
functions. Very recently, Tetzlaff et al. (2010) carried out a comprehensive 
study on Hipparcos stars, and built a catalogue of runaway stars. Following 
former ideas, the authors fit spatial velocity distributions to the 
objects found\footnote{We will not consider here the so-called {\it 
hypervelocity stars}, which are originated in three-body encounters that 
include a high mass black hole like the one in the Galactic center.}.\\

\begin{figure}[!ht]
  \centering
  \includegraphics[width=.7\textwidth]{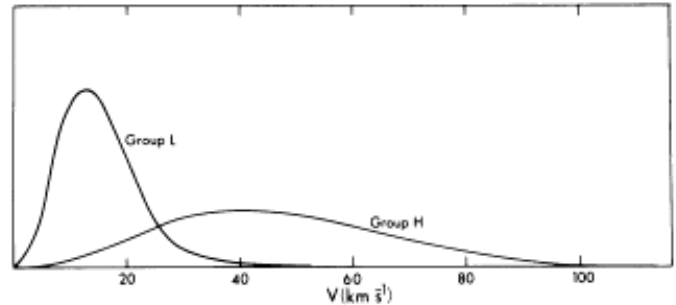}~\hfill
  \caption{Velocity frequency functions for the two groups of O stars  
up to 2 kpc; L: low velocity stars, H: high velocity stars (Stone 1979).}
  \label{stone-groups}
\end{figure}

The input sample of Tetzlaff et al. (2010) consisted of a database of thousands 
of stars up to 3 kpc from the Sun. By computing runaway star probabilities, the 
authors catalogued a 27\% of runaways throughout the sample. The importance of 
this study lies on the extent and uniformity of the sample, which is not biased 
by spectral type or visual magnitude, like previous ones.

Conclusive evidences that at least two different mechanisms operate in nature 
to `kick' a star have been presented (e.g., Hoogerwerf et al. 2000). The 
processes are referred as the Binary Supernova Scenario (BSS) and the Dynamical 
Ejection Scenario (DES). In the last years, extensive N-body simulations have 
been carried out, that could explain the observed ejection rate solely by the 
action of close gravitational encounters (see Perets \& Subr 2013 and 
references therein). The knowledge of the exact multiplicity status 
of the runaway star 
is crucial to test its origin\footnote{As Virpi Niemela likes to say: ''with 
evidence at hand, it is ussually straightforward to state that a star is a 
binary, but it is more than difficult to ensure that is single''.}. More 
realistic results will be obtained as long as forthcoming instruments allow 
deeper studies.

Runaway stars can be used as tools to trace galactic structure. Silva \& 
Napiwotzki (2013) have shown, with a local sample of high-latitude runaway 
stars, that an analysis of the birthplaces help to map the spiral arms and 
determine galactic dynamics. The method will be terribly powerful when 
instruments like {\it Gaia} come to play.

\section{Stellar bow shocks: theory and observations}

A star moving through a slower interstellar medium will form a thin 
layer of swept-up gas. An early-type star, that has developed strong stellar 
winds (OB or Wolf-Rayet star) will stack matter on a thicker layer.

If the stellar motion is supersonic with respect to the ambient gas velocity, 
shock waves are produced. A discontinuity surface is formed, and two shock 
fronts in opposite directions. A `forward' shock in the direction of the 
stellar motion travels with a velocity similar to the stellar velocity. A 
`reverse' shock from the discontinuity to the star has a velocity of the order 
of the stellar wind terminal velocity, which can be, in some cases, a few 
thousands of km s$^{-1}$. In Figure 2 we show a simplified scheme of the 
situation. The width of the discontinuity will depend on the cooling process: 
adiabatic, radiative or a mixed case. Ideally, four regions can be identified: 
the closer to the star with free wind, that of shocked wind (where the reverse 
shocked has passed), the shocked ambient matter region and the 
not-yet-disturbed ISM region.

\begin{figure}[!h]
  \centering
  \includegraphics[width=.7\textwidth]{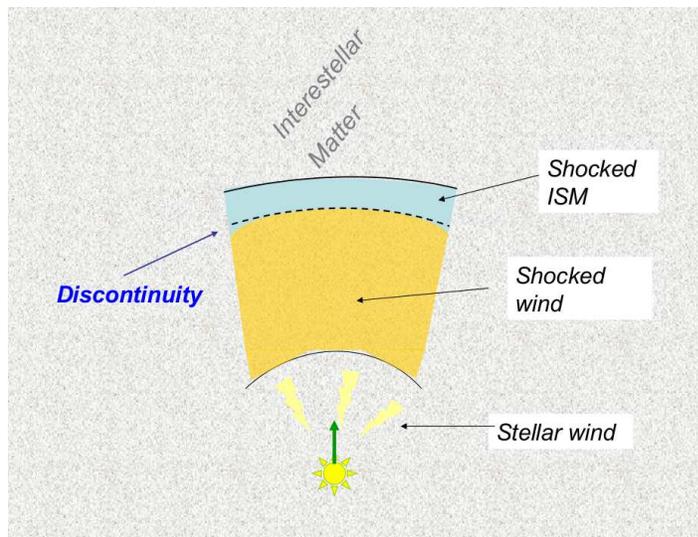}~\hfill
  \caption{Scheme of a stellar bow shock generated by a star with supersonic
motion related to the ambient media. The different regimes are shown. 
The green vector represents the stellar velocity.}
  \label{fig2}
\end{figure}

The matter surrounding the stars is piled-up in a feature that resembles the 
sea foam which is pushed by the bow of a ship, from where the stellar bow 
shock takes its name. For the faster stars, it resembles a cometary tail.

The arc-shaped structure moves in the same direction of the stellar velocity 
vector. The stellar winds are confined by the ISM pressure. The point where the 
momentum of the wind balances that of the ISM is defined as the stagnation 
point $R_0$.

The shocked ISM heats the swept-up matter, and the dust re-radiates
% amorphous silicate dust or graphitic carbon dus.
strongly as an excess at MID and FIR wavelengths. Consequently, stellar bow 
shocks of massive, early type stars are revealed at infrared frequencies.
% the radius of IS dust grains is about microns, and T is about 20K.
In Figure 3 we present the infrared emission from a stellar bow shock produced 
by the runaway star HD 77581, better known as Vela X-1, a HMXB formed by a B0.5 
Iae star (Prinja \& Massa 2010) and a compact object. The IR images are part of 
the Wide-field Infrared Survey Explorer (WISE, Wright et al. 2010). The 
structure is detected at the observed bands 1 to 4 (respectively centred at 
3.4, 4.6, 12, and 
22 $\mu$m). Since the heat can be enough to ionise the gas, some bow shocks are 
also seen in the H$\alpha$ emission line (like this one, Kaper et al. 1997) and 
in radio continuum. The dust contains policyclic aromatic hydrocarbons (PAHs) with 
emission features at some WISE bands (1 and 3). The features are excited by 
non- ionizing UV photons from the stellar radiation field and the dust becomes 
bright. The stellar UV field also ionizes the interstellar gas, which 
de-excites via recombination lines like Br$\alpha$ (WISE band 2). The WISE band 
4 is sensitive to emission from warm dust.

From the IR images it is possible to derive the density of the ISM, by 
measuring the distance from the star to the shock, once the stellar velocity, 
the mass loss rate and the wind terminal velocity are known.

\begin{figure}[!h]
  \centering
  \includegraphics[width=0.24\textwidth]{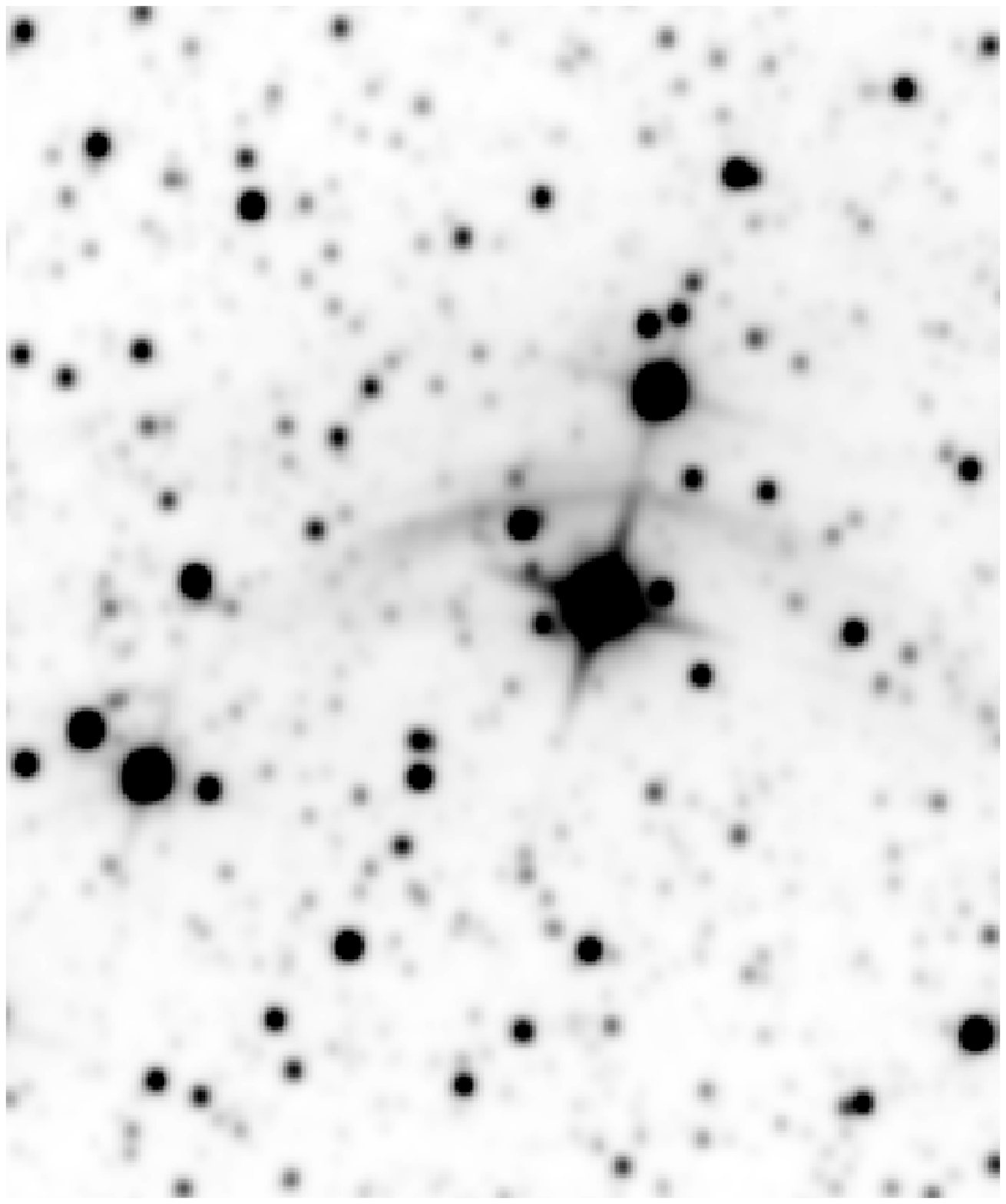}
  \includegraphics[width=0.24\textwidth]{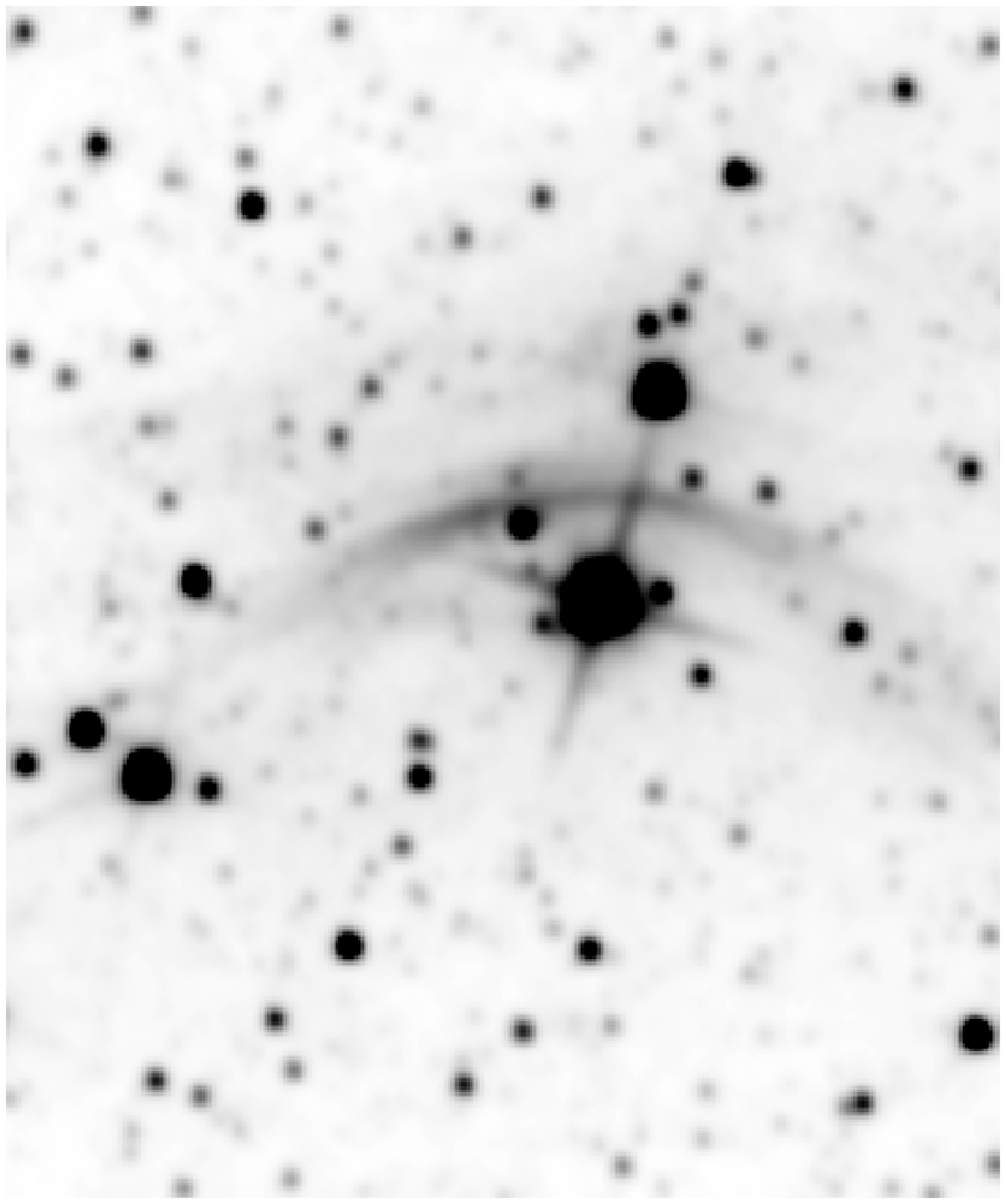}
  \includegraphics[width=0.24\textwidth]{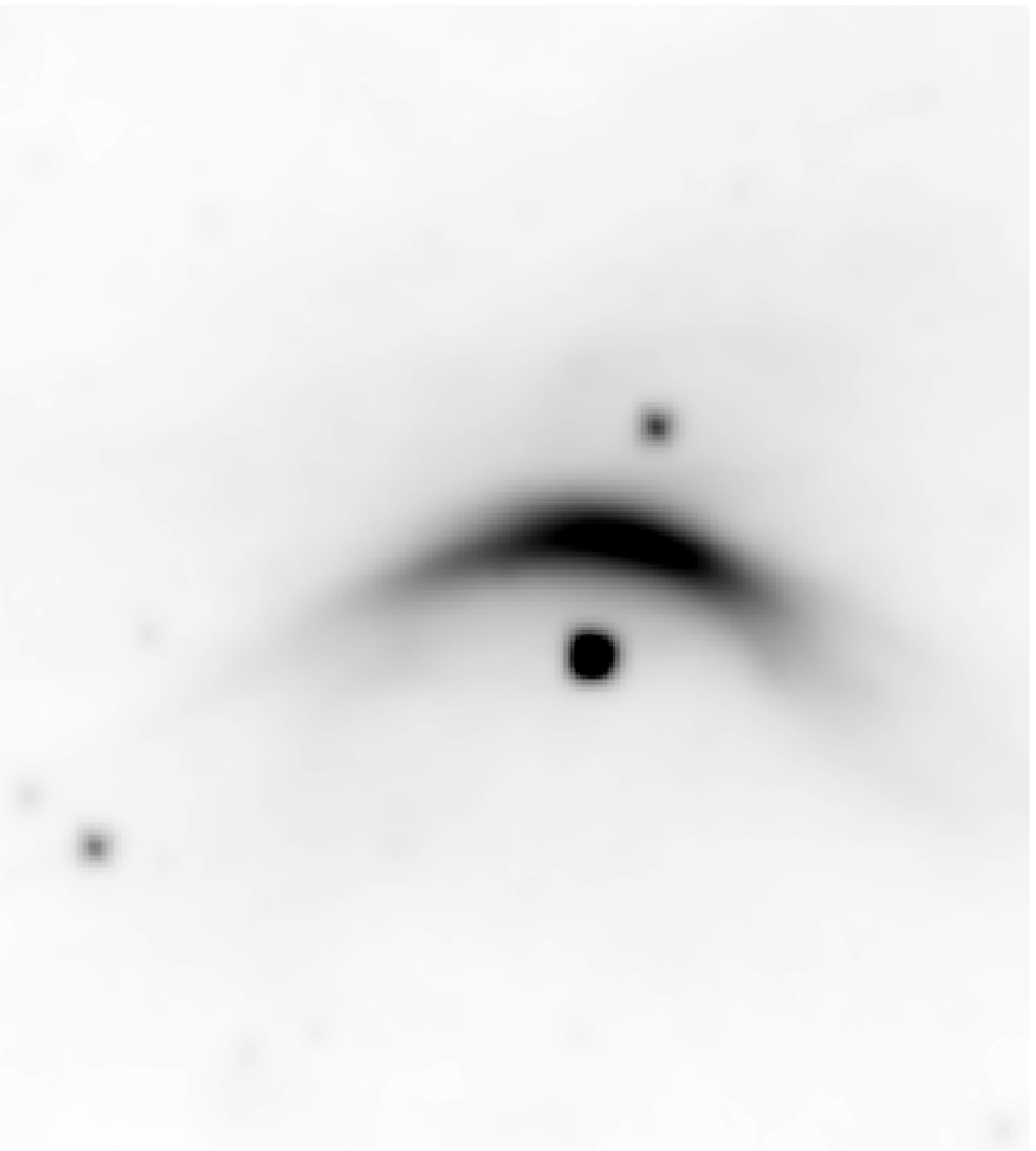}
  \includegraphics[width=0.24\textwidth]{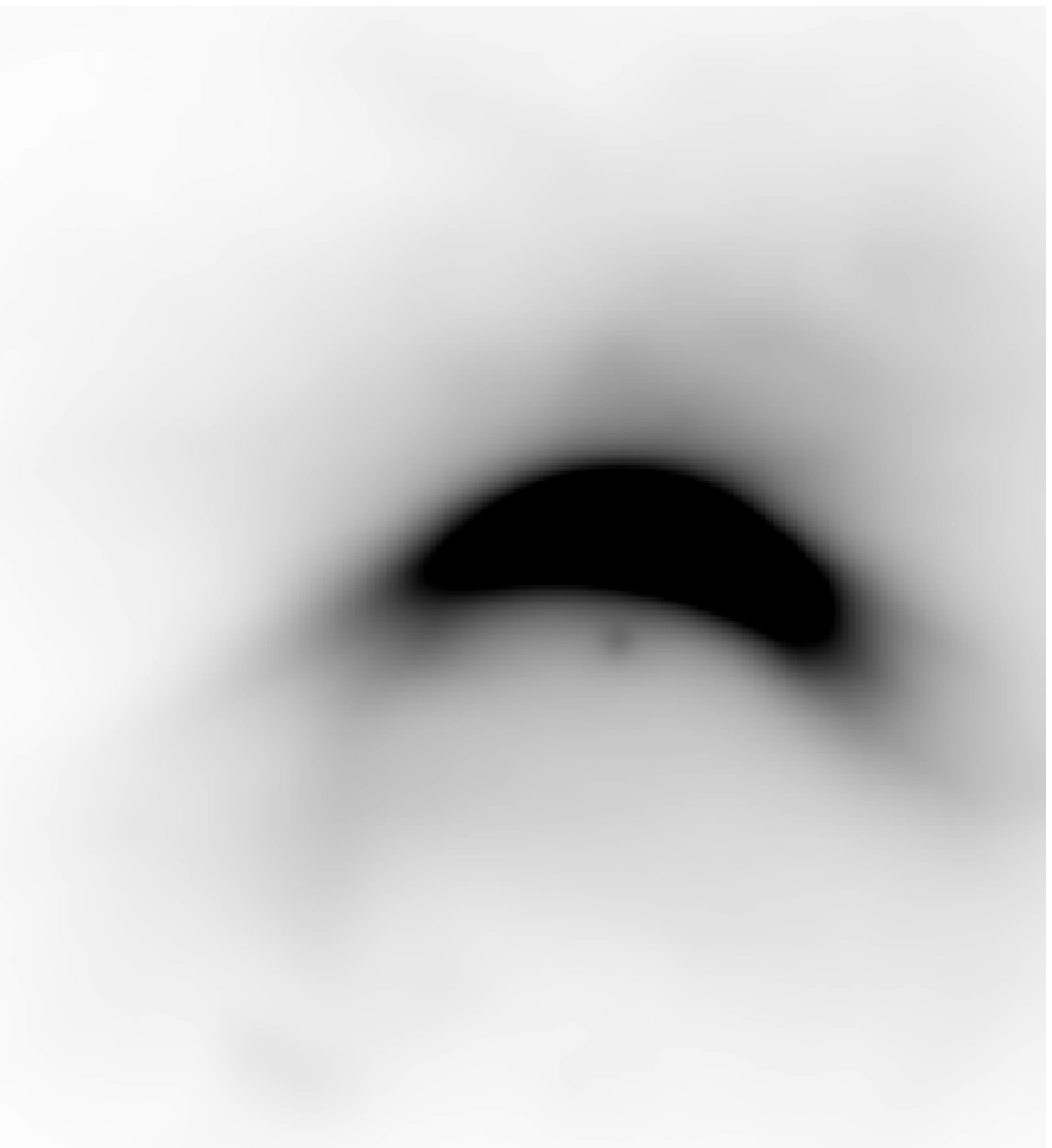}
  \caption{Field of HD 77581 (Vela X-1): emission in WISE bands 1 
(3.4 $\mu$m), 2 (4.6 $\mu$m), 3 (12 $\mu$m) and 4  (22 $\mu$m), from left 
to right. The bow shock feature and the star HD 77581 
are seen at all bands. The star can be better identified at band 3, 
close to the bow shock and to the south.}
  \label{bowshock-ir}
\end{figure}

There is much theoretical work carried out on bow shock dynamics (see 
references in Peri et al. 2012, and del Valle and Romero 2012, for instance). 
The 
first attempts considered 1-D modelling, a thin-shell, uniform density. 
Gradually, they included density gradients, misaligned winds, slow and fast 
winds, fluid instabilities, clumped winds, (non-uniform) magnetic fields, 
rotating flows, etc. To cite a few, Gustafsson et al. (2010) built 3D models of 
interstellar bow shocks propagating in a homogeneous molecular medium with a 
uniform magnetid field. The authors found that the bow shock shape depends 
strongly on the orientation of the magnetic field, and could reproduce
H$_2$ emission lines detected from the bow shock of the source OMC1.

Cox et al. (2012) carried out 2D hydrodynamical simulations of interactions 
between the ISM and the circumstellar medium to analyze how the morphology of 
the bow shock varies with stellar wind and ISM parameters. They used a model to 
describe bow shocks around AGB stars as seen by data obtained with the Herschel 
space telescope.

Acreman, Stevens and Harries (2013) made use of hydrodynamical modelling and 
Monte-Carlo radiative transfer calculations to find the radiation influenced 
region around a runaway early-type star. The authors generated IR, H$\alpha$ 
and radio observables and obtained a good match between synthetic and real 
images.

Despite detailed studies on individual 'famous' stars with bow shocks, 
or clusters 
rich in early-type stars (i.e. Vela X-1: Kaper et al. 1997; Cyg OB2: Kobulnicky 
et al. 2010; BD+43$^\circ$ 3654: Comer\'on \& Pasquali 1998, to quote a few), 
one can still 
ask basic questions like in which scenarios (ISM and star) a bow shock is 
formed, how it evolves or under which conditions it is detected. A crucial 
issue is how observations fill in theoretical developments or, more precisely, 
how models fit the data? Undoubtely, the study of a large-enough sample of such 
objects will bring the answers closer, and allow to perform real statistics.

\section{Stellar bow shock surveys}

After the outstanding view of the sky by the IRAS satellite, Noriega-Crespo et 
al. (1997) analized the infrared emission from the surrounding field of about 
60 early-type stars. The angular resolution of the data was $\sim 1'$. The 
emission was classified as diffuse, bubble- or bow-shock like. The authors 
presented results of about 20 stellar bow shock candidates among the total 
sample.

Following IR space telescopes improved the data angular resolution, like the 
case of the Midcourse Space eXperiment (MSX, 18$''$, Egan et al. 2003). The 
IRAS (all-sky covering) successor was the Wide-field Infrared Survey 
Explorer\footnote{Funded by NASA; contributing institutions: 
UCLA, JPL, IPAC/Caltech, UC Berkeley, SDL, BATC.}. WISE observing bands  
resolutions were 6.1, 6.4, 6.5, 12$''$ respectively. The sensitivity 
resulted in hundreds of times better than that of IRAS. The telescope 
was active from December 2009 to February 2011. In early 2011 the WISE 
team published a first data release encompassing observations towards
57\% of the sky.

At the same time, a large database of runaway stars was made publicly 
available: the catalogue of young runaway Hipparcos stars within 3 kpc from the 
Sun already mentioned (Tetzlaff et al. 2010). 
From an initial sample of 7663 stars, the 
authors compiled information on proper motions, parallaxes, spectral types, 
radial velocities and estimated distances, ages, spatial velocites, and 
aggregate membership. After a probabilistic study, by means of mainly computed 
velocities, they converged on a catalogue of runaway stars candidates of about 
2400 objects.

Taking into account two kinds of samples, the one of Noriega-Crespo et al. 
(1997), and the Tetzlaff et al. (2011) catalogue, Peri et al. (2012) conducted 
a systematic search of stellar bow shocks. 
The first list had already detected IR stellar bow shocks, so the main 
goal was to compare IRAS data with WISE data. In many cases the 
classification of stellar bow shock remained, and also new bow shocks were 
discovered. For the second sample, the authors seeked to test the theoretical 
assumption that the runaway stars generate observable stellar bows shocks. For 
the first list, that of Noriega-Crespo et al., they found 14 of the IRAS bow 
shocks in WISE, and 4 in addition through WISE and MSX data; the rest had no 
new IR data, or no bow shock.
For the second sample (only stars of spectral 
types from O to B2, in total 244), they found 17 bow shocks, 80 objects with no 
data on WISE, and 147 with no bow shock. Results on the first list confirm the 
already observed bow shocks, and brought some new examples. Those of the second 
list showed that about a 10\% of the sources had an associated bow shocks. But 
probably the most important result was the confirmation that the stellar bow 
shocks can form ot not, and if they form they can have different structures. 
This conclusion can be analized looking at the IR images and the several 
parameters that caracterize all the stellar bow shock candidates. This issue 
has been studied through numerical simulations by Comer\'on and Kaper (1998) 
and it 
seems so far that the observations correlate with the existing models very 
well. 

E-BOSS bow shocks are depicted in Figure 4. On March 2012 the WISE team 
published IR images and data of the remaining sky (43\%). The second version
of E-BOSS is under way, and it will include results on the search at the WISE
second release database.

\begin{figure}
  \centering
  \includegraphics[width=0.9\textwidth]{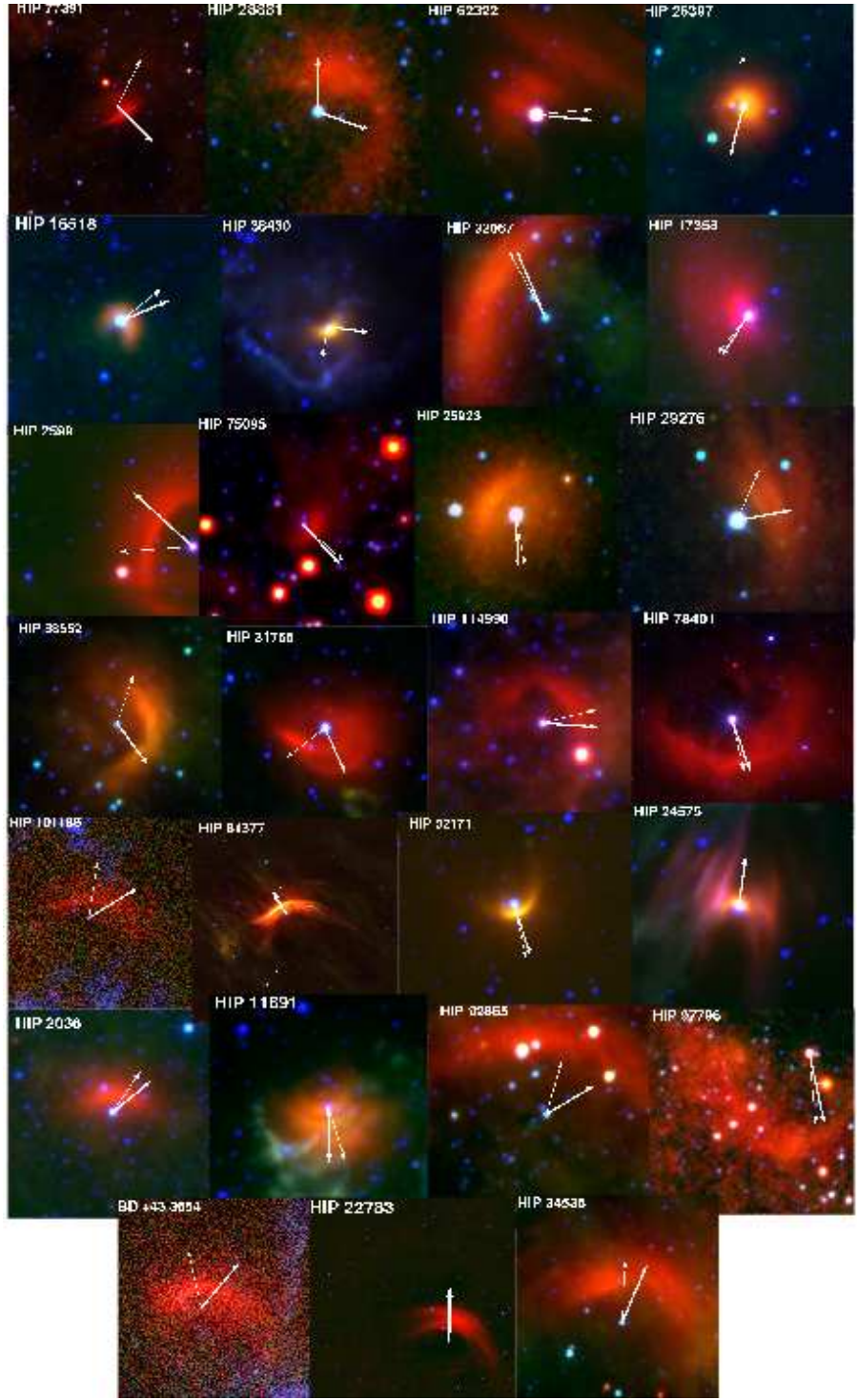}
  \caption{WISE images of the E-BOSS.v1 members (Peri et al. 2012). 
%Color mapping: {\bf blue}=3.4 microns; {\bf green}=12.1 microns;
%{\bf red}=22.2 microns. The color scales are in data numbers (DNs);
%the WISE data is not calibrated in their surface brightness.
Vectors: proper motions from Hipparcos (thicker) and corrected for the 
ISM Galactic rotation. The star name is given in the top right corner.}
  \label{eboss}
\end{figure}

\section{Bow shocks as particle acceleration regions}

Nearby windy stars with supersonic motion relative to the ISM, particles can be 
accelerated up to relativistic velocities via the first order Fermi mechanism by 
repeatedly crossing the discontinuity surface (Bell 1978). It can be 
demonstrated that protons easily diffuse towards the 'tail' of the bow shock, 
but that is not the case for the -less-massive- electrons.

The shock wave compresses the gas. The magnetic field is coupled with the gas, 
and increases its energy density. Consequently, the magnetic field is much 
amplified at the post-shock region, where the shock wave is stronger ($\propto$ 
wind terminal velocity) (e.g. Longair 1997), to $\sim 10^{-5}$ Gauss. 
The interaction of the relativistic electrons with the post-shock $B$ field 
will give rise to synchrotron radiation, detectable at radio wavelengths. In 
Figure 5 we represent the scenario: the ISM magnetic field $B_{\rm ISM}$, the 
amplified field at the post-shocked region ($B'$), the electrons gaining energy 
by crossing the discontinuity surface, and the protons diffusing outwardly.

\begin{figure}
  \centering
  \includegraphics[width=0.6\textwidth]{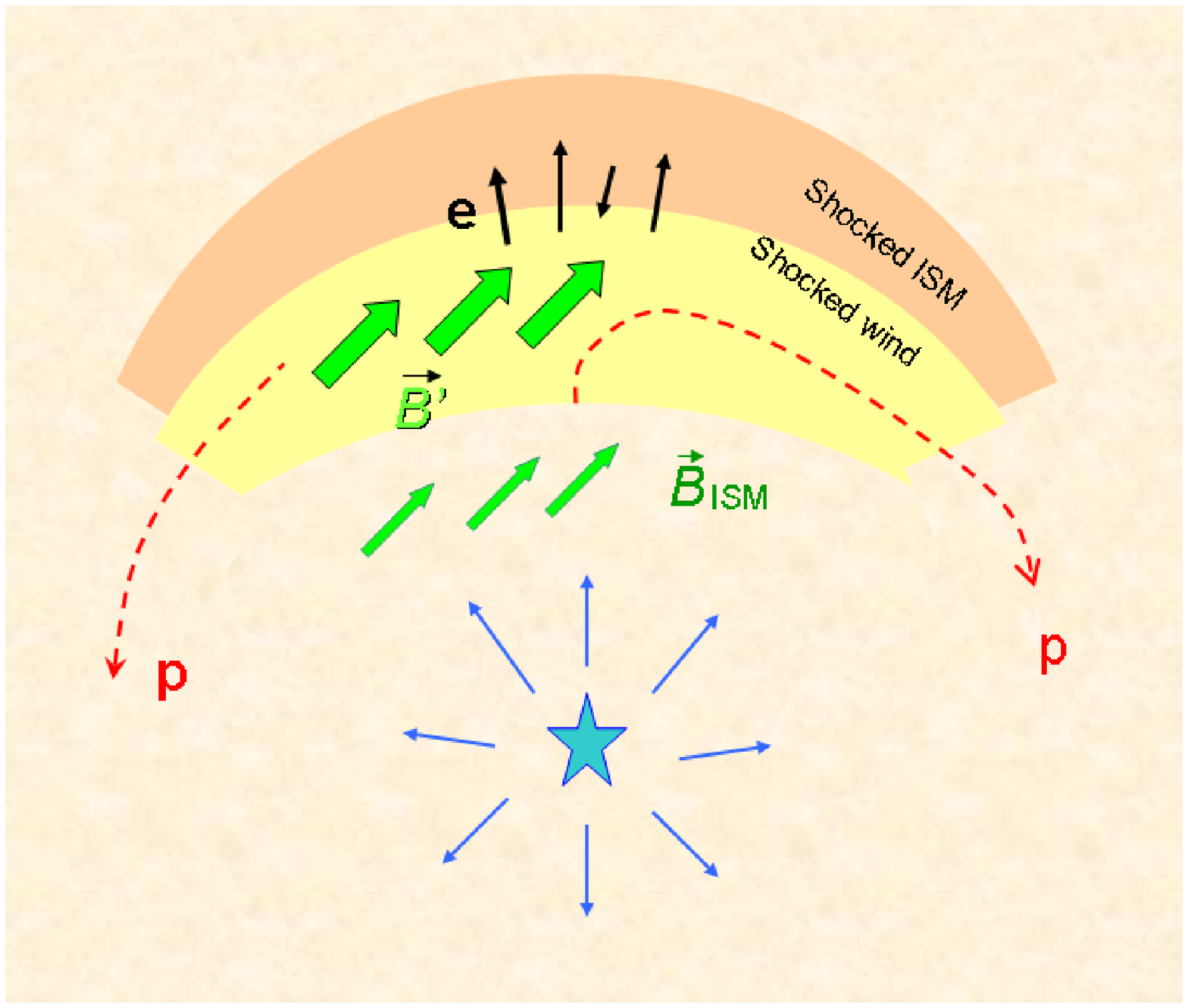}
  \caption{Regions of particle acceleration up to relativistic energies: the 
star (in blue), the free stellar wind (blue arrows); the shocked wind region 
(light yellow), the shocked ISM region (pale orange). 'p' is for protons, and 
'e' for electrons. The interstellar magnetic field ${\bf B}_{\rm ISM}$ is 
represented by the smaller green arrows. This magnetic field is amplified at 
the post-shocked region, as ${\bf B}'$.}
  \label{eboss}
\end{figure}

To look for signatures of synchrotron emission we carried out radio 
interferometric observations toward the field of the O4 supergiant BD 
+43$^{\circ}$ 3654 (Benaglia et al. 2010). Comer\'on \& Pasquali (2007) had 
proposed that this is a runaway star from the Cyg OB2 association, and found a 
bow shock feature at MSX images. The radio data were taken at two bands: 1.4 
and 4.8 GHz, and the bow shock was detected at both of them. 
A spectral index map and 
its corresponding error map were then built (see Figure 6). The average value 
of the spectral index resulted in $\sim -0.4$, consistent with non-thermal 
emission. Conclusive evidence of pure synchrotron emission will confirm the 
presence of a relativistic electrons population. These particles will also be 
involved in high energy processes.

Later observations with the Giant Metrewave Radio Telescope at 0.61 and 1.25 
GHz by Brookes et al. (2013) confirmed the former results.

\begin{figure}[!h]
  \centering
  \includegraphics[width=0.28\textwidth,angle=-90]{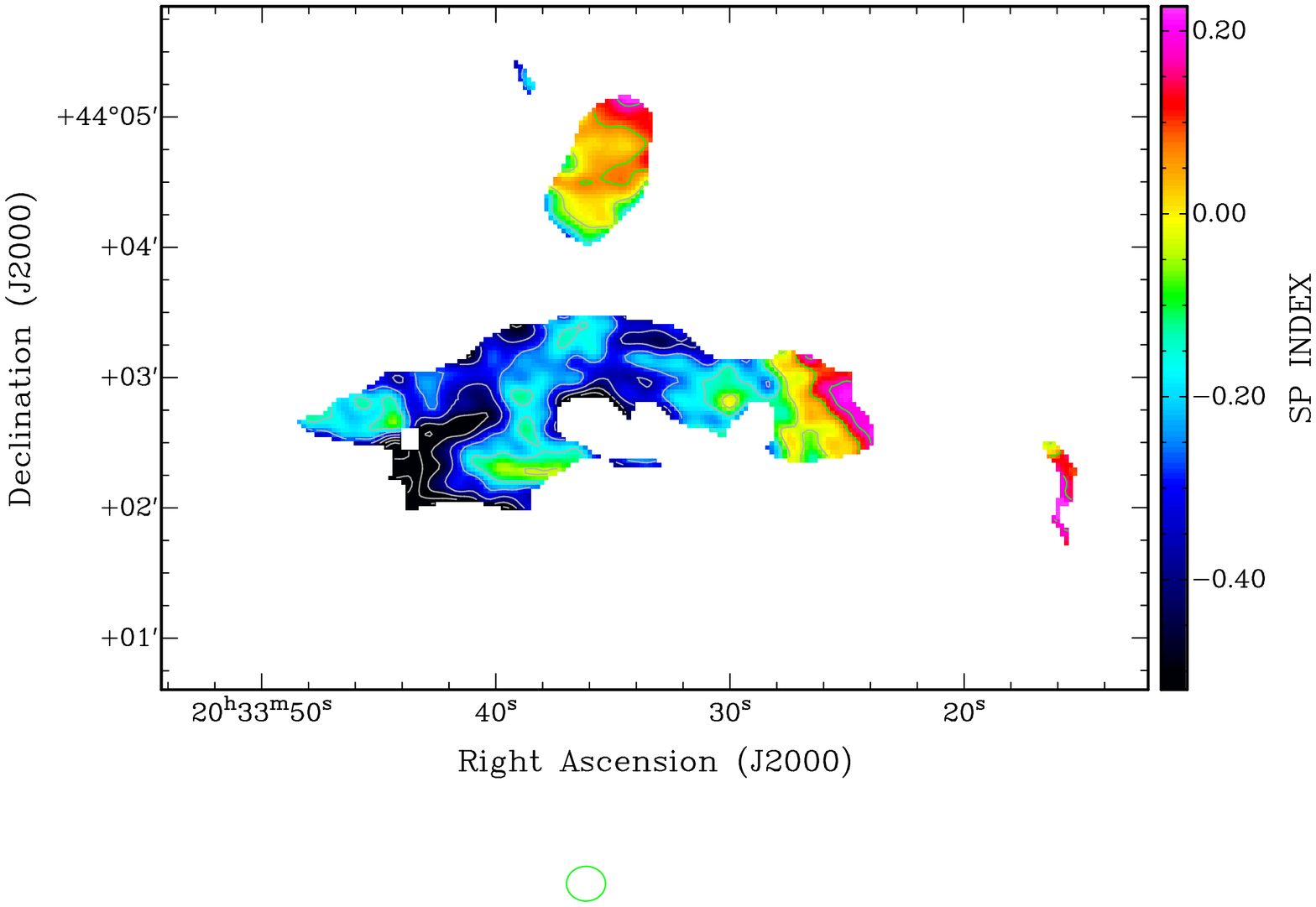}
  \includegraphics[width=0.28\textwidth,angle=-90]{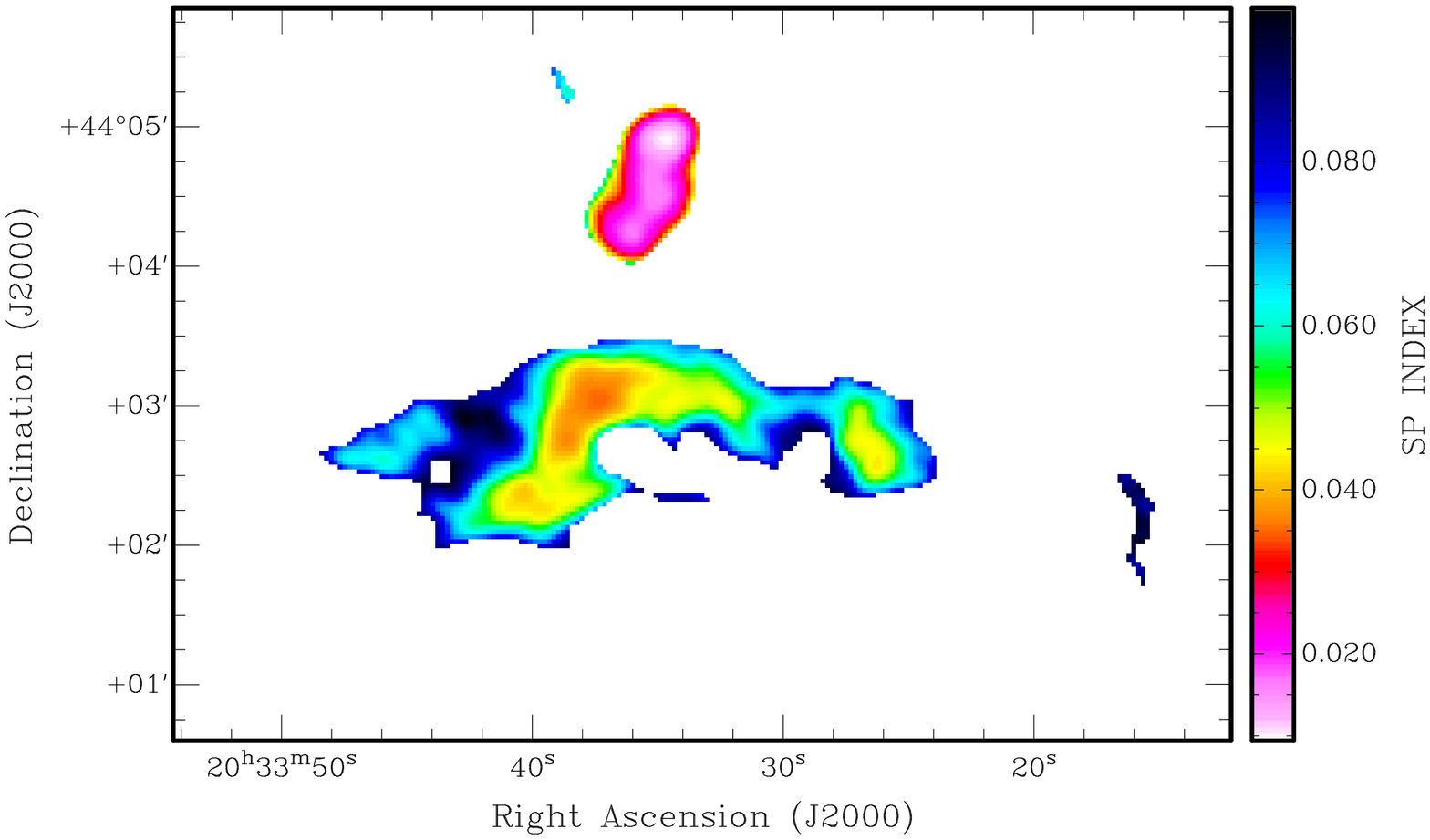}
 \caption{Spectral index map and spectral index error map of the 
radio emission associated with the stellar bow shock of BD+43$^0$3654 (see
Benaglia et al. 2010).}
  \label{spixes-bd}
\end{figure}

If the radio emission is mainly produced by synchrotron processes, one can use 
the measured radio flux and spectral index as input for a model of the SED. 
Del Valle 
\& Romero (2012) showed that the bow shock of a star like $\zeta$ Oph, detected 
at WISE bands, could produce, under certain conditions, high energy emission 
measurable by forthcoming instruments like the Cerenkov Telescope Array (see 
Figure 7).

\begin{figure}[!ht]
  \centering
  \includegraphics[width=0.88\textwidth]{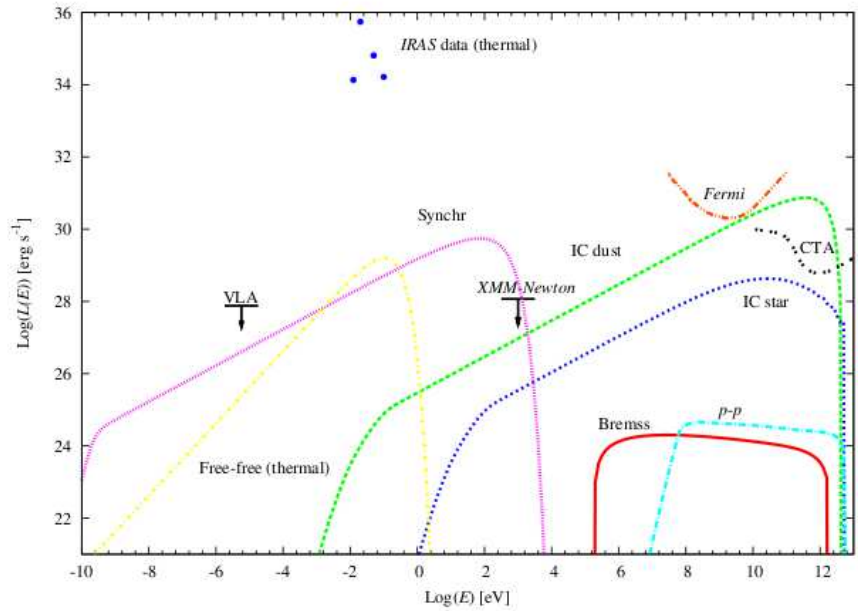}
 \caption{Spectral energy distribution of $\zeta$ Oph (see del Valle \& Romero 
2012). Various mechanisms for high energy production have been modelled.}
  \label{zetaoph}
\end{figure}

Another way to look for non-thermal emission is by means of X-ray data 
analysis. Lopez-Santiago et al. (2012) studied the fields of the E-BOSS 
candidates through XMM-Newton observations. They detected the star HIP 24575 
(AE Aur), but also an XMM source, BS, on the position of the bow shock (see 
Figure 8). They showed that the BS emission could be fitted by a non-thermal 
component.

\begin{figure}[!ht]
  \centering
  \includegraphics[width=0.48\textwidth]{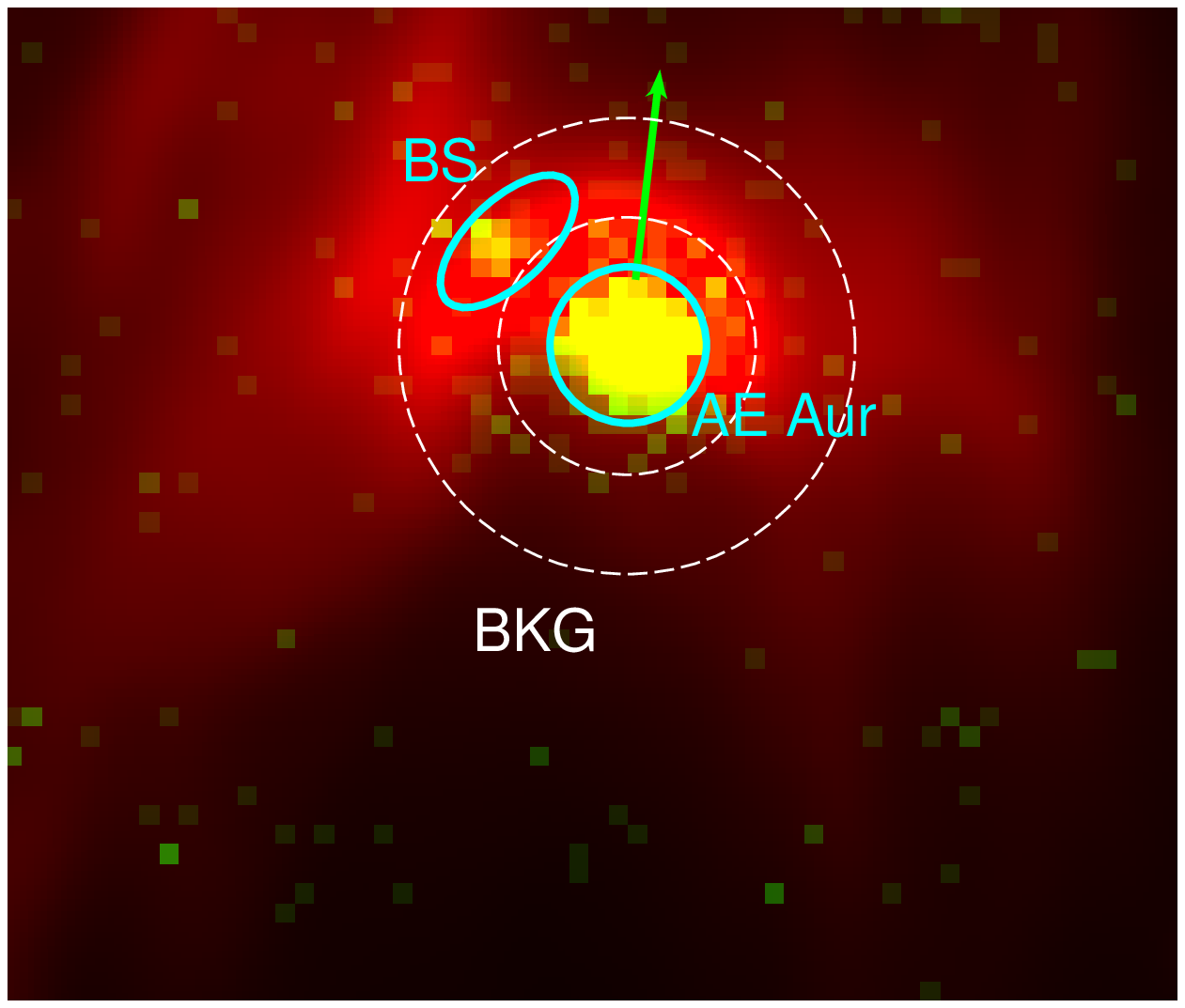}
  \includegraphics[width=0.48\textwidth]{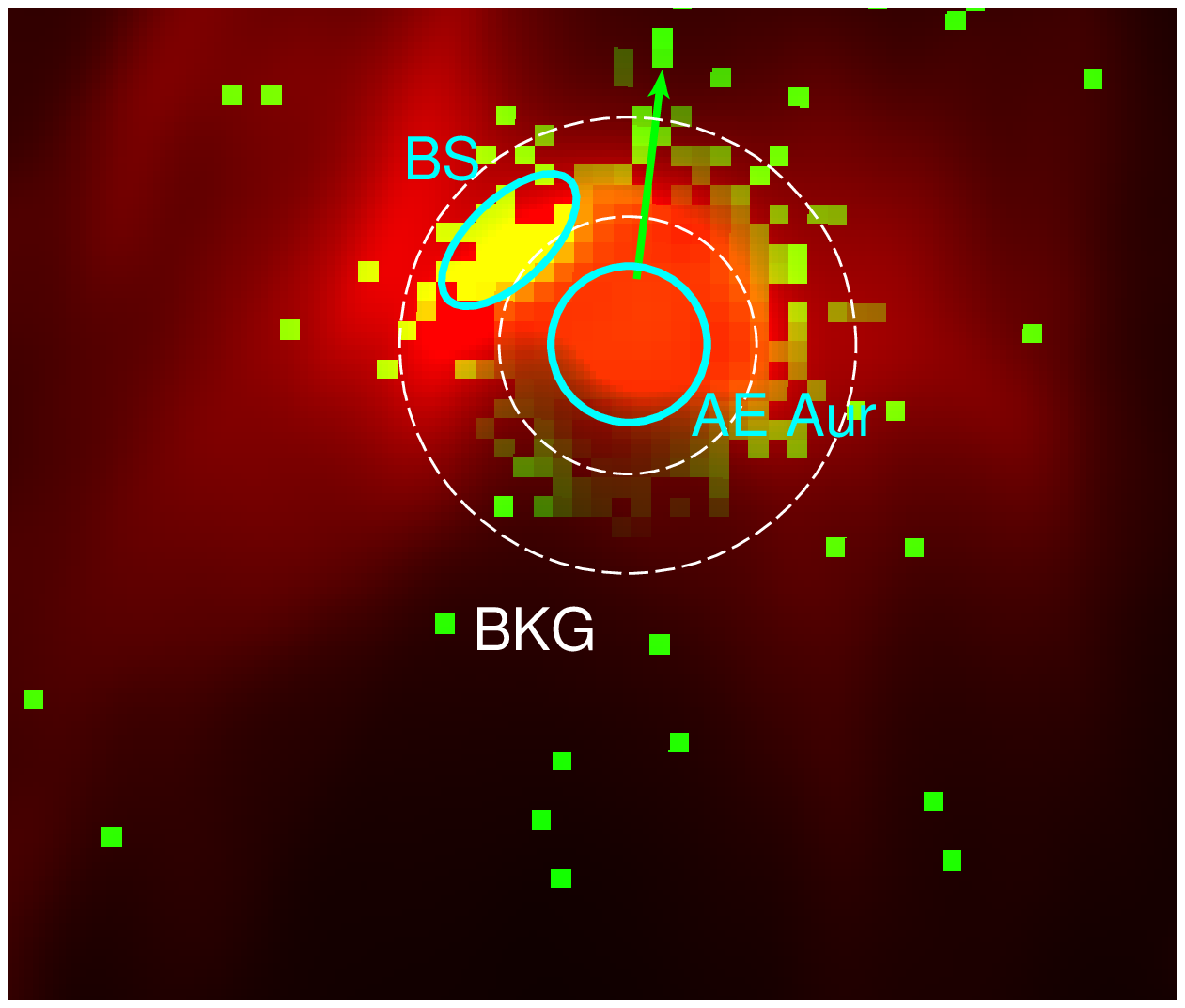}
 \caption{In red: WISE $12.1 \mu$m image. In yellow: XMM-Newton EPIC 
emission in the keV bands $1-8$ (right), and $0.3-8$ (left).}
  \label{aeaurigad}
\end{figure}

%\section{Some statistics}

\section{Further studies}

Detailed studies on stellar bow shocks, either planned or under way, 
include the followings:

\begin{itemize}

\item Carry out bow shock searches around stars with spectral types different 
from O-B2;

\item Perform statistics over the full (final) bow shock candidate catalogue;

\item Improve 3D modelling adding magnetohydrodynamics, thermal radio emission, 
and wind and ISM inhomogeneities;

\item Implement dedicated observations toward the bow shocks with signs of 
non-thermal radio emission;

\item Look for polarised radio emission from bow shocks to confirm synchrotron origin.

\end{itemize}

\acknowledgments P.B. acknowledges the LOC and SOC of the Annual Meeting of the 
Asociaci\'on Argentina de Astrnom\'{\i}a. PB and CSP were partially supported 
by FONCyT, PICT 00848.

\subsection*{References}

\noindent Acreman, D.M., Stevens, I.R., Harries, T.J. 2013, in press

\noindent Bell, A.R. 1978, MNRAS 182, 443

%\bibitem[2010]{benaglia} 
\noindent Benaglia, P., Romero, G.E., Mart\'i, J., Peri, C.S., Araudo, A.T. 2010, A\&A, 517, L10

%\bibitem[1961]{blaauw} 

\noindent Blaauw, A. \& Morgan 1954, ApJ, 119, 625

\noindent Blauuw, A. 1961, BAN, 15, 265

\noindent Brookes, D., et al. 2013, in preparation

\noindent L\'opez-Santiago, J., Micheli, M., del Valle, M.V., et al. 2012, ApJ, 757

%\bibitem[2007]{comeron} 
\noindent Comer\'on, F. \& Pasquali, A. 2007, A\&A, 467, L23

\noindent Cox, N.L.J., Kerschbaum, F., van Marle, A.J., et al. 2012, A\&A, 543,1

\noindent del Valle, M.V. \& Romero, G.E. 2012, A\&A, 543, 56

%\noindent del Valle, M.V., Romero, G.E., De Becker, M. 2013, A\&A, 550, 112

\noindent Egan, M.P., Price, S.D., Kraemer, K.E. 2003, AAS, 203, 5708

\noindent Gustafsson, M., Ravkilde, T., Kristensen, et al. 2010, A\&A 513, 5

%\bibitem[2000]{hoog-2000} 
\noindent Hoogerwerf, R., de Bruijne, J.H.J, de Zeeuw, P.T. 2000, ApJ, 544, L133

%\bibitem[1997]{kaper} 
\noindent Kaper, L., Van Loon, J.Th., Augusteijn, T. et al. 1997, ApJ, 479, L153

%\bibitem[2010]{kobulnicky} 
\noindent Kobulnicky, H.A., Gilbert, I.J., Kiminki, D.C. 2010, ApJ, 710, 549

\noindent Longair, M. 1997, ``High Energy Astrophysics'', Cambridge University Press

\noindent Noriega-Crespo, A., Van Buren, D., Dgani, R. 1997, AJ, 113, 780

\noindent Perets, H.B. \& Subr, L. 2012, ApJ, 751, 133

\noindent Peri, C.S., Benaglia, P., Brookes, D.P., et al. 2012, A\&A, 538, 108

\noindent Prinja, R.K. \& Massa, D.L. 2010, A\&A, 521, 55

\noindent  Silva, M.D.V. \& Napiwotzki, R. 2013, MNRAS, in press

\noindent Stone, R.C. 1979, ApJ, 232, 520

%\noindent Terada, Y., Tashiro, M., Bamba, A., et al. 2012, PASJ, 64, 138

\noindent Tetzlaff, N., Neuh\"auser, R., Hohle, M.M. 2010, MNRAS, 410, 190

\noindent Wright, E.L., Eisenhardt, P.R.M., Mainzer, A.K. et al. 2010, AJ, 140, 6, 1868

\noindent Zwicky, F. 1957, {\sl Morphological Astronomy}, SpringerVerlag, Berlin

%\bibliographystyle{baaa}
%\bibliography{Benaglia-IIbiblio}

\end{document}